\documentstyle[12pt]{article}
\def\appendix#1{
\addtocounter{section}{1}
\setcounter{equation}{0}
\renewcommand{\thesection}{\Alph{section}}
\section*{Appendix \thesection\protect\indent #1}
\addcontentsline{toc}{section}{Appendix \thesection\ \ \ #1}
}

\def\be{\begin{equation}}
\def\la{\label}
\def\ee{\end{equation}}
\def\bea{\begin{eqnarray}}
\def\eea{\end{eqnarray}}

\begin{document}
\title{On Hamiltonian structure of the spin Ruijsenaars-Schneider 
model}
\author{G.E.Arutyunov
\thanks{Steklov Mathematical Institute,
Gubkina 8, GSP-1, 117966, Moscow, Russia; arut@genesis.mi.ras.ru}\\
and \\
S.A.Frolov
\thanks{Steklov Mathematical Institute,
Gubkina 8, GSP-1, 117966, Moscow, Russia; frolov@genesis.mi.ras.ru}\\
}
\date {}
\maketitle
\begin{abstract}
The Hamiltonian structure of spin generalization of the 
rational Ruijsenaars-Schneider model is found 
by using the Hamiltonian reduction 
technique. It is shown that the model possesses the current algebra 
symmetry. The possibility of generalizing
the found Poisson structure to the trigonometric
case is discussed and  degeneration to the 
Euler-Calogero-Moser system is examined.  
\end{abstract}

\section{Introduction}
Recently a spin generalization \cite{KrZ} of the elliptic 
Ruijsenaars-Schneider model \cite{RS,R} (spin RS model) was introduced as 
a dynamical system describing the pole evolution of the elliptic 
solutions of the non-abelian 2D Toda chain. Equations of motion 
proposed for the model generalize the ones for the  
Euler-Calogero-Moser system (ECM) \cite{GH}-\cite{KBBT}, which is an 
integrable system of $N$ particles with internal degrees of freedom 
interacting by a special pairwise potential. 

An important tool for dealing with classical integrable systems and 
especially for quantizing them is the Hamiltonian formalism.
Although equations of motion defining the spin RS 
model can be integrated in terms of Riemann theta-functions, 
the question about their Hamiltonian form remains open. 
The aim of the present paper is to give a partial answer
to this question, which lies in constructing the explicit Hamiltonian 
formulation for the rational spin RS model.         

Our construction is based on the Hamiltonian reduction procedure
acknowledged as the unifying approach to dynamical systems of 
Calogero or Ruijsenaars type \cite{KKS}-\cite{ACF}. In this approach one 
starts with a large initial phase space and a simple Hamiltonian possessing 
a symmetry group. Then factorizing the corresponding motion by this 
symmetry one is left with a nontrivial dynamical system defined on a 
reduced phase space. In particular, the rational RS model and 
the trigonometric Calogero-Moser system appear in this way if one uses the 
cotangent bundle $T^{*}G$ over a Lie group $G$ as the initial phase space 
\cite{GNe}. 

A natural generalization of this approach allowing us to include 
spin variables consists in replacing $T^{*}G$ by a more general
phase space $\cal P$ that we choose to be $T^{*}G\times {\cal J}^*$, where 
${\cal J}^*$ is a dual space to the Lie algebra ${\cal J}$ of $G$.
Considering on $\cal P$ a special Hamiltonian $H_R$ and
performing the Hamiltonian reduction by $G$-action,
we obtain the Poisson structure of the rational spin RS model.

Let us briefly describe the content of the paper and the  
results obtained. For simplicity, we restrict ourselves to the case of 
$G=GL(N,{\bf C})$. In Section 2 we define on $\cal P$ two 
dynamical systems governed by Hamiltonians $H_C$ and $H_R$ and show 
that the corresponding integrals of motion combine into generators of the 
Yangian and the current algebra respectively. Since all these
integrals are gauge-invariant, the corresponding symmetries will 
survive after the reduction. 

As is known \cite{Ne} the dynamical system on the 
reduced phase space corresponding to $H_C$ is the trigonometric ECM model.
This immediately reproduces the result found in \cite{BGHP,BAB} that the 
model possesses the Yangian symmetry.

Section 3 is devoted to the rational spin RS model. First we introduce
$G$-invariant spin variables that after solving the moment map equation
can be identified with coordinates on the reduced phase space 
${\cal P}_r$. Equations of motion for dynamical variables of 
${\cal P}_r$ produced by $H_R$ coincide with the ones introduced in 
\cite{KrZ} for the rational case. That is the way we obtain an explicit 
Hamiltonian formulation of the spin RS model. The Poisson structure of the 
model is found to be rather nontrivial and admits at least two equivalent 
descriptions in terms of different phase variables. Moreover, it depends 
on a parameter $\gamma$ being a coupling constant of the model.

It turns out that the spin RS model admits such 
(spectral-independent) $L$-operator (Lax matrix) that satisfies the same 
$L$-operator algebra as for the corresponding spinless model.
We also show that the Hamiltonian reduction
provides an alternative way of solving equations of motion without
using spectral curves. A similar method of integrating equations of
motion of the spin RS model was used in \cite{RaS}.

Finally, we present an explicit expression for generators 
of the current algebra via phase variables of the spin RS model and 
define the gauge-invariant  momentum variables.

In Section 4 degeneration of the rational spin RS model to the ECM system 
is examined. An interesting feature we come here is the appearance of 
spin variables obeying the defining relations of the Frobenius Lie 
algebra. We observe that the general elliptic ECM system can 
be also formulated in terms of Frobenius spin variables.

\section{Current and Yangian symmetries}
In this section we construct representations of the Yangian and current
algebras related to the cotangent bundle $T^*G$ over the matrix group 
$G=GL(N,{\bf C})$ and describe their connection to the ECM and
the spin RS models respectively.
 
Consider the following manifold ${\cal P}=T^*G\times {\cal G}^*$,
where ${\cal G}^*$ is a dual space to the Lie algebra
${\cal G}=\mbox{Mat}(N,{\bf C})$ of $G$. We parametrize an element
from ${\cal G}^*$ by a matrix $S\in {\cal G}$ due to the
isomorphism ${\cal G}^{*}\approx {\cal G}$. 
The space $T^*G$ is naturally isomorphic to ${\cal G}^*\times G$
and we parametrize it by pairs $(A,g)$, where 
$A\in {\cal G}$ and $g\in G$. The algebra of
regular functions on ${\cal P}$
is supplied with a Poisson structure, which can be written
in terms of variables $(A,g,S)$ as follows 
\begin{eqnarray} 
\{ A_1,A_2\} &=&\frac 1 2 [C,A_1 
-A_2] \label{AA}\\
\{ A_1,g_2\} 
&=&g_2 C,~~~~\{g_1,g_2\}=0, \label{pb} \\
\{S_1,g_2\}&=&\{S_1,A_2\}=0
\label{gg} \\
\{ S_1,S_2\} &=&-\frac 1 2 [C,S_1 -S_2]. 
\label{SS}
\end{eqnarray} 
Here we use the standard tensor notation 
and $C=\sum_{i,j}E_{ij}\otimes E_{ji}$ is the permutation operator.  

The Poisson structure is invariant under the following action
of the group $G$:
\be
A\to hAh^{-1},~~~g\to hgh^{-1},~~~S\to hSh^{-1}.
\la{sym}
\ee
We refer to (\ref{sym}) as to gauge transformations.
The moment map of this action is of the form
\be
\mu=gAg^{-1}-A+S.
\la{mom}
\ee
The simplest gauge-invariant Hamiltonians  
are $H_C=\mbox{tr}A^2$ and $H_R=\mbox{tr}g$. 

The Poisson bracket of the variables $S_{ij}$ can be realized
by using  $2N$ $l$-dimensional vectors $a_i,b_i$ 
which form $lN$-pairs of canonically conjugated variables:
$$
\{a^{\alpha}_i,b^{\beta}_j\}=-\delta_{ij}\delta^{\alpha\beta},
$$
where $i,j=1,\cdots ,N$ and $\alpha ,\beta=1,\cdots , l$.
Supposing the matrix elements of $S$ to be
\be
S_{ij}=\sum_{\alpha}a^{\alpha}_ib^{\alpha}_j
\la{Ss}
\ee
one recovers the Poisson bracket (\ref{SS}). 
Obviously, under gauge transformations the variables $a$ and $b$
transform in the following way
$$
a^{\alpha}\to ha^{\alpha},~~~b^{\alpha}\to b^{\alpha}h^{-1},
$$
where we regard $a^{\alpha}$ as a column and $b^{\alpha}$ as a row.

The variables
$a$ and $b$ allow one to construct a lot of gauge-invariants 
Poisson commuting with $H_C$ or with $H_R$.

First we consider a family of integrals of motion for $H_C$:
$I^{\alpha\beta}_n=\mbox{tr}A^nS^{\alpha\beta}$, where for any $\alpha$
and $\beta$ the matrix $S^{\alpha\beta}$ has the entries 
$S^{\alpha\beta}_{ij}=a^{\alpha}_ib^{\beta}_j$. In fact, the integrals
$I_n$ form a representation of the classical Yangian. To see this,
one can introduce the following generating function 
$T^{\alpha\beta}(z)$ of $I_n$:
$$
T^{\alpha\beta}(z)=\delta^{\alpha\beta}+
\mbox{tr}\frac{1}{z-A}S^{\alpha\beta}.
$$
By using the Poisson bracket for the variables $S^{\alpha\beta}$:
$$
\{S_1^{\alpha\beta},S_2^{\mu\nu}\}=
C_{12}(\delta^{\beta\mu}S_2^{\alpha\nu}-\delta^{\alpha\nu}S_1^{\mu\beta})
$$
and performing simple calculations, one obtains the Yangian algebra
\be
\{T_1(z),T_2(w)\}=[r(z-w),T_1(z)T_2(w)].
\la{Y}
\ee
Here we regard $T(z)$ as an $l\times l$-matrix with entries 
$T^{\alpha\beta}(z)$, and $r(z-w)$ is a rational solution of
the classical Yang-Baxter equation:
$$
r(z-w)=\frac{K}{z-w},
$$
where $K$ is the permutation operator acting in 
${\bf C}^l\otimes {\bf C}^l$. 

A well-known property of the Yangian is the existence of the
involutive subalgebra generated by $I_k(z)=\mbox{tr}T(z)^k$.

For the Hamiltonian  $H_R$ one can choose the following family of 
integrals of motion $J^{\alpha\beta}_n=\mbox{tr}g^nS^{\alpha\beta}$.
Introducing the formal generating function $J(z)$:
$$
J^{\alpha\beta}(z)=\sum_{n=-\infty}^{\infty}J^{\alpha\beta}_nz^{-n-1},
$$
one can easily show that $J(z)$ satisfies the current algebra relations:
\be
\{J_1(z),J_2(w)\}=[K,J_2(w)]\delta(\frac{z}{w}),
\la{J}
\ee
where $\delta(\frac{z}{w})=\frac{1}{z}
\sum_{n=-\infty}^{\infty}(\frac{z}{w})^{n}$ is the formal 
$\delta$-function. 

It is obvious that $\mbox{tr}J(z)^n$ are central elements of 
the current algebra. In addition, the current algebra admits
an involutive family of integrals of motion polynomial in $g$ and $S$.
It is constructed as $J_n^{+}(z)=\mbox{tr}J^+(z)^n$, where 
$J^+(z)=\sum_{n=0}^{\infty}J_nz^{-n-1}$. The involutivity is a 
consequence of the algebra satisfied by $J^+(z)$: 
\be
\{J^+_1(z),J^+_2(w)\}=[r(z-w),J^+_1(z)+J^+_2(w)].
\la{J+}
\ee
It is well known that the Yangian (\ref{Y}) is a deformation of 
(\ref{J+}).

The dynamical systems governed by the Hamiltonians $H_C$ and $H_R$  
are trivial. However, factorizing the initial phase space $\cal P$
by the action of a symmetry group, one gets nontrivial systems
defined on the reduced phase space ${\cal P}_r$. In particular,
to get the ECM and the spin RS models one should fix the moment
map as \cite{GN}
\be
gAg^{-1}-A+S=\gamma I,
\la{momf}
\ee
where $\gamma$ is a complex number being identified with 
a coupling constant. Then solving this equation modulo the action
of the gauge group $G$ ($G$ coincides with the isotropy group of 
(\ref{momf}), i.e. (\ref{momf}) is a set of first-class constraints), 
one obtains the reduced phase space. The dynamical
systems on ${\cal P}_r$ corresponding to  the Hamiltonians $H_C$ and
$H_R$ are identified with the ECM and the spin RS models respectively.

Since the generators of the Yangian and current algebras are
gauge-invariant, we conclude that the ECM model possesses the Yangian
symmetry whereas the spin RS model has the current symmetry. 
As was mentioned in the Introduction 
this result for the ECM model was obtained in \cite{BGHP,BAB} by exploiting
an explicit $L$-operator describing the model. 

\section{Rational spin RS model}
In this section we present the Hamiltonian formulation of the 
spin RS model by considering the reduction of the phase space $\cal P$
by the action of $G$. Given the moment map, the space of
functions on 
the reduced phase space ${\cal P}_{r}$ can be identified with the space 
$\mbox{Fun}^G{\cal P}$ of $G$-invariant functions
on $\cal P$ restricted to the surface (\ref{momf}) of the constant moment 
level. A choice of an appropriate basis in $\mbox{Fun}^G{\cal P}$ and 
calculation of the induced Poisson structure make the description of 
${\cal P}_{r}$ explicit. 

To construct a basis in $\mbox{Fun}^G{\cal P}$ we first note
that any semisimple element of $\cal G$ can be diagonalized 
by a gauge transformation:
\be
A=TQT^{-1},
\la{d}
\ee
where $Q$ is a diagonal matrix with gauge-invariant entries 
$q_i\neq q_j$. By using the action of the Weyl group we fix the order of 
$q_i$. For given $A$, the matrix $T$ in (\ref{d}) is uniquely defined 
by requiring to be an element of the Frobenius group, i.e. it satisfies
the condition \be Te=e, \la{f} \ee where $e$ is an $N$-dimensional 
column with all $e_i=1$. Such a choice for $T$ is known \cite{AF} to 
be relevant for the description of the RS model. 

Given $A$ and $g$, we can diagonalize a matrix $A'=gAg^{-1}=UQU^{-1}$
with the help of an element $U$ such that $Ue=e$. This introduces a useful
parametrization for $g$: $g=UPT^{-1}$, where $P$ is some diagonal
matrix.

Under gauge transformation (\ref{sym}) matrices $T$ and $U$ transform as
follows $T\to hTh[T]$, $U\to hUh[U]$ where $h[T]$, $h[U]$ are 
diagonal matrices $h[T]_i=(T^{-1}h^{-1}e)_i$, $h[U]_i=(U^{-1}h^{-1}e)_i$. 

Introduce diagonal matrices 
$t_{ij}=t_i\delta_{ij}$ and $u_{ij}=u_i\delta_{ij}$  
with entries
\be
t_i=\sum_{\alpha}(T^{-1}a^{\alpha})_i,~~~
u_i=\sum_{\alpha}(U^{-1}a^{\alpha})_i
\la{sp}
\ee
which transform under gauge transformation (\ref{sym})  in
the following way 
$$ t_i\to h[T]_i^{-1}t_i,~~~
u_i\to h[U]_i^{-1}u_i. 
$$ 
We use $t$ to define $G$-invariant spin variables 
$$ 
{\bf a}^{\alpha}_i=t_i^{-1}(T^{-1}a^{\alpha})_i,~~~ 
{\bf c}^{\alpha}_i=t_i(b^{\alpha}UP)_i.
$$ 
Note that ${\bf a}^{\alpha}_i$ are not arbitrary but satisfy 
constraints $\sum_{\alpha}{\bf a}^{\alpha}_i=1$ for any $i$.
The relevance of this definition will be clarified later.

To calculate the Poisson algebra of ${\bf a}$ and ${\bf c}$
one needs to use the one for $(T,U,P,Q)$-variables.
In \cite{AF} it was proved that the standard Poisson structure 
(\ref{AA},\ref{pb})
on $T^*G$ rewritten in terms of $(T,U,P,Q)$-variables has the form
\bea
\la{TT}
\{T_1,T_2\}&=&T_1T_2r_{12},~~~\{U_1,U_2\}=-U_1U_2r_{12}, \\
\la{TP}
\{T_1,P_2\}&=&T_1P_2\bar{r}_{12},~~~\{U_1,P_2\}=U_1P_2\bar{r}_{12}, \\
\la{TQ}
\{T_1,Q_2\}&=&\{U_1,Q_2\}=\{P_1,P_2\}=\{T_1,U_2\}=0, \\
\la{QP}
\{Q_1,Q_2\}&=&0,~~~\{Q_1,P_2\}=P_2\sum_i E_{ii}\otimes E_{ii}.
\eea
Here $r_{12}$ is an $N$-parametric solution of the classical Yang-Baxter
equation:
\be
r_{12}=\sum_{i\neq j}\frac{1}{q_{ij}}F_{ij}\otimes F_{ji},
\la{r}
\ee
where $F_{ij}=E_{ii}-E_{ij}$ is a basis of the Frobenius Lie algebra 
and the matrix $\bar{r}_{12}$ is  given by 
\be
\bar{r}_{12}=\sum_{i\neq j}\frac{1}{q_{ij}}F_{ij}\otimes E_{jj}.
\la{rch}
\ee
Note that (\ref{QP}) implies that $q_i$ and $p_i=\log P_i$ are
canonically conjugated variables.

With formulae (\ref{TT})-(\ref{QP}) at hand we first calculate 
$$
\{t_i,t_j\}=-\frac{1}{q_{ij}}(t_i-t_j)^2,~~
\{u_i,u_j\}=\frac{1}{q_{ij}}(u_i-u_j)^2,~~\{t_i,u_j\}=0
$$
and then the Poisson brackets of the invariant spins:
\bea
\la{aa}
\{{\bf a}^{\alpha}_i,{\bf a}^{\beta}_j\}&=&
\frac{1}{q_{ij}}
({\bf a}^{\alpha}_i{\bf a}^{\beta}_j+{\bf a}^{\alpha}_j{\bf a}^{\beta}_i
-{\bf a}^{\alpha}_i{\bf a}^{\beta}_i-{\bf a}^{\alpha}_j{\bf a}^{\beta}_j) 
,\\
\la{cc}
\{{\bf c}^{\alpha}_i,{\bf c}^{\beta}_j\}&=&
\frac{1}{q_{ij}}
({\bf c}^{\alpha}_i{\bf c}^{\beta}_j+{\bf c}^{\alpha}_j{\bf 
c}^{\beta}_i)+
{\bf c}^{\beta}_j {\bf L}_{ji}-{\bf L}_{ij}{\bf c}^{\alpha}_i,
\\
\la{ca}
\{{\bf c}^{\alpha}_i,{\bf a}^{\beta}_j\}&=& 
\delta^{\alpha\beta}{\bf L}_{ji}-
{\bf a}^{\beta}_j{\bf L}_{ji}
+\frac{1}{q_{ij}}{\bf c}^{\alpha}_i({\bf a}^{\beta}_i
-{\bf a}^{\beta}_j),\\
\la{ac}
\{{\bf a}^{\alpha}_i,{\bf c}^{\beta}_j\}&=& 
-\delta^{\alpha\beta}{\bf L}_{ij}+
{\bf a}^{\alpha}_i{\bf L}_{ij}
+\frac{1}{q_{ij}}{\bf c}^{\beta}_j({\bf a}^{\alpha}_j
-{\bf a}^{\alpha}_i).
\eea 
The Poisson structure of invariant spins is not closed since it involves 
another gauge invariant object ${\bf L}$:  
${\bf  L}_{ij}=t_i^{-1}L_{ij}t_j$, where $L=T^{-1}gT$.  
In \cite{AF} $L$ was identified with the $L$-operator of the rational
RS model. Analogously, ${\bf L}$ will be called the $L$-operator of the 
spin RS model. The relevance of this definition will be justified  later.
Note that in eqs.(\ref{aa})-(\ref{ac}) and in formulas below
it is assumed that if some denominator becomes zero,
the corresponding fraction is omitted. 

Calculating the Poisson algebra for ${\bf L}$ with the
help of eqs.(\ref{TT})-(\ref{QP}), we obtain that it coincides with the
one for $L$, namely
\bea
\la{L}
\{{\bf L}_1,{\bf L}_2\}&=&r_{12}{\bf L}_1{\bf L}_2+{\bf L}_1{\bf L}_2 
\hat{r}_{12} 
+{\bf L}_1\bar{r}_{21}{\bf L}_2-{\bf L}_2\bar{r}_{12}{\bf L}_1,
\eea
where $\hat{r}_{12}=\bar{r}_{12}-\bar{r}_{21}-r_{12}$
is a constant solution of the Gervais-Neveu-Felder equation \cite{GN,Fel}.
Therefore, the $L$-operator algebra for the spin RS model and the one 
for the RS model without spins are the same. 

To complete the description of the Poisson algebra of invariant spins
we find the Poisson brackets of ${\bf L}$ with ${\bf a}$ and ${\bf c}$:
\bea
\la{aL}
\{{\bf a}^{\alpha}_i,{\bf L}_{kl}\}&=&
\frac{1}{q_{ik}}({\bf a}^{\alpha}_i-{\bf a}^{\alpha}_k){\bf L}_{kl}-
\frac{1}{q_{ik}}({\bf a}^{\alpha}_i-{\bf a}^{\alpha}_k){\bf L}_{il}-
\frac{1}{q_{il}}({\bf a}^{\alpha}_i-{\bf a}^{\alpha}_l){\bf L}_{kl},
~~~~~\\
\la{cL}
\{{\bf c}^{\alpha}_i,{\bf L}_{kl}\}&=&
\frac{1}{q_{il}}{\bf c}^{\alpha}_i{\bf L}_{kl}+
\frac{1}{q_{il}}{\bf c}^{\alpha}_l{\bf L}_{ki}-
\frac{1}{q_{ik}}{\bf c}^{\alpha}_i{\bf L}_{kl}+
\frac{1}{q_{ik}}{\bf c}^{\alpha}_i{\bf L}_{il}\\
\nonumber
&+&{\bf L}_{li}{\bf L}_{kl}-{\bf L}_{ki}{\bf L}_{kl}.
\eea
Thus, the Poisson algebra of gauge-invariant variables ${\bf a},{\bf c}$
and ${\bf L}$ is closed. 

Remark that the choice of gauge-invariant spins
and the $L$-operator is not unique. In particular, one could use
$\omega_i=\sum_{\alpha}(b^{\alpha}T)_i$ to define other gauge-invariant
spins $\hat{a}^{\alpha}_i=\omega_i (T^{-1}a^{\alpha})_i$, 
$\hat{c}^{\alpha}_i=\omega_i^{-1} (b^{\alpha}UP)_i$  and
the $L$-operator: $\hat{L}_{ij}=\omega_iL_{ij}\omega_j^{-1}$. One can
verify that this set of gauge-invariant variables satisfies a 
different algebra.

The next step consists in restricting the Poisson algebra 
(\ref{aa})-(\ref{cL}) to the surface (\ref{momf}) of the constant moment 
level. Diagonalizing the variable $A$, we find that
eq.(\ref{momf}) is equivalent to 
\be
LQ-QL-\gamma L = - T^{-1}STL.
\la{red}
\ee
Multiplying (\ref{red}) by the diagonal matrix $t$ from the right
and by $t^{-1}$ from the left, and taking into account that 
$L=T^{-1}gT=T^{-1}UP$, we rewrite (\ref{red}) in terms
of gauge-invariant variables
\be
{\bf L}Q-Q{\bf L}-\gamma {\bf L} = - t^{-1}T^{-1}SUPt,
\la{red1}
\ee
since 
$(t^{-1}T^{-1}SUPt)_{ij}=
\sum_{\alpha}{\bf a}^{\alpha}_i{\bf c}^{\alpha}_j$.
Thus, we can solve eq.(\ref{red1}) with respect to ${\bf L}$. The
solution is given by 
\be
{\bf L}=\sum_{ij}\frac{f_{ij}}{q_{ij}+\gamma}E_{ij},
\la{Li}
\ee
where we introduced 
$f_{ij}=\sum_{\alpha}{\bf a}^{\alpha}_i{\bf c}^{\alpha}_j$. 
Now the reduction of the Poisson structure (\ref{aa})-(\ref{cL})
on the surface (\ref{red}) amounts to the substitution in the r.h.s. 
of (\ref{aa})-(\ref{cL}) the entries of the $L$-operator (\ref{Li}). 
The consistency of the reduced Poisson structure 
can be also checked by direct calculations.
To this end one first find the Poisson algebra of $f_{ij}$-variables:
\bea
\nonumber
\{f_{ij},f_{kl}\}&=&
(\frac{1}{q_{ik}}+\frac{1}{q_{jl}}+\frac{1}{q_{kj}}+\frac{1}{q_{li}})f_{ij}f_{kl}
+(\frac{1}{q_{ki}}+\frac{1}{q_{il}+\gamma})f_{ij}f_{il}\\
\nonumber
&+&
(\frac{1}{q_{ik}}+\frac{1}{q_{jl}}+\frac{1}{q_{kj}+\gamma}-\frac{1}{q_{il}+\gamma})f_{il}f_{kj}
+(\frac{1}{q_{jk}}-\frac{1}{q_{jl}+\gamma})f_{ij}f_{jl}\\
&+&
(\frac{1}{q_{ki}}-\frac{1}{q_{kj}+\gamma})f_{kj}f_{kl}+
(\frac{1}{q_{il}}+\frac{1}{q_{lj}+\gamma})f_{lj}f_{kl}
\la{ff}
\eea
and $\{f_{ij},q_k\}=-f_{ij}\delta_{kj}$. Then using the representation
(\ref{Li}) for ${\bf L}$ one does recover the $L$-operator algebra (\ref{L}).

Now we proceed with describing dynamics on ${\cal P}_r$. The invariant
Hamiltonian $H_R$ acquires on ${\cal P}_r$ a form 
$H_R=\mbox{tr}{\bf L}=\frac{1}{\gamma}\sum_{i}f_{ii}$. 
This Hamiltonian and the Poisson 
structure on ${\cal P}_r$ produce the following equations of motion 
\bea 
\la{qdd}
&&\dot{q}_i=L_{ii}=\frac{1}{\gamma}f_{ii},\\
\la{ad}
&&\dot{{\bf a}}^{\alpha}_i=-\sum_{j\neq i}
\frac{1}{q_{ij}}({\bf a}^{\alpha}_i-{\bf a}^{\alpha}_j) {\bf L}_{ij}=
-\frac{1}{\gamma}
\sum_{j\neq i}({\bf a}^{\alpha}_i-{\bf a}^{\alpha}_j)f_{ij}V(q_{ij}),
\\
\la{cd}
&&\dot{{\bf c}}^{\alpha}_i=\sum_{j\neq i}
\frac{1}{q_{ij}}({\bf c}^{\alpha}_i{\bf L}_{ij}+
{\bf c}^{\alpha}_j {\bf L}_{ji})=\frac{1}{\gamma}
\sum_{j\neq i}({\bf c}^{\alpha}_if_{ij}V(q_{ij})-
{\bf c}^{\alpha}_jf_{ji}V(q_{ji})),~~~
\eea
where we introduced the potential 
$V(q_{ij})=\frac{1}{q_{ij}}-\frac{1}{q_{ij}+\gamma}$. Differentiating
$\dot{q}_i$ and taking into account eqs.(\ref{ad}),(\ref{cd}),
one gets
\be
\la{qd}
\ddot{q}_i=2\sum_{j\neq i}\frac{1}{q_{ij}}L_{ij}L_{ji}=
2\sum_{j\neq i}\frac{1}{q_{ij}}{\bf L}_{ij}{\bf L}_{ji}=
\frac{1}{\gamma^2}\sum_{j\neq i}f_{ij}f_{ji}(V(q_{ij})-V(q_{ji})),
\ee
and equations of motion for $f_{ij}$:
\be
\la{eqf}
\dot{f_{ij}}=\frac{1}{\gamma}\sum_{k\neq i,j}
V(q_{kj})f_{ik}f_{kj}-V(q_{ik})f_{ik}f_{kj}+V(q_{ik})f_{ik}f_{ij}-
V(q_{jk})f_{jk}f_{ij}.
\ee
It follows from eqs.(\ref{qdd}) and (\ref{eqf}) that the equation of 
motion for ${\bf L}$ can be written in the Lax form 
$\dot{{\bf L}}=[{\bf L},{\bf M}]$ with ${\bf M}=\sum_{i\neq 
j}\frac{1}{q_{ij}}{\bf L}_{ji}F_{ij}$. However, this equation is
not equivalent to eqs.(\ref{qdd}) and (\ref{eqf}). 

In the paper \cite{KrZ} the 
spin generalization of the elliptic RS model was introduced. The 
generalized model is a system of $N$ particles with coordinates 
$q_i$, each particle has internal degrees of freedom described by 
$l$-dimensional vector $a^{\alpha}_i$ and $l$-dimensional vector 
$c^{\alpha}_i$. The equations of motion generalize the ones for the 
ECM system:  \bea \la{qsK} \ddot{q}_i&=&\sum_{j\neq 
i}f_{ij}f_{ji}(V(q_{ij})-V(q_{ji})),\\ 
\la{aK} 
\dot{a}_i&=&\sum_{j\neq 
i}{a}_j f_{ij}V(q_{ij})-\lambda_ia_i,\\ 
\la{bK} 
\dot{c}_i&=&-\sum_{j\neq 
i}{c}_j f_{ji}V(q_{ji})+\lambda_ic_i, 
\eea 
where 
$V(q)=\zeta(q)-\zeta(q+\gamma)$, $\lambda_i(t)$ are arbitrary 
functions of $t$ and 
$\zeta(q)$ denotes the Weierstrass zeta-function. Equations of motion 
(\ref{qsK}-\ref{bK}) are invariant under rescaling:
$$
a_i\to k_ia_i,~~~c_i\to \frac{1}{k_i}c_i.
$$
Introducing the invariant variables
$\hat{a}_i^{\alpha}=(\sum_{\alpha}a_i^{\alpha})^{-1}a_i^{\alpha}$ and
$\hat{c}_i^{\alpha}=(\sum_{\alpha}a_i^{\alpha})c_i^{\alpha}$, and
calculating from (\ref{aK}) and (\ref{bK}) equations of motion for
$\hat{a}_i^{\alpha}$ and $\hat{c}_i^{\alpha}$, one discovers that 
all $\lambda_i$ drop out and equations of motion coincide with (\ref{ad})
and (\ref{cd}) with the change $\hat{a}^{\alpha}_i\to {\bf a}_i^{\alpha}$, 
$\hat{c}^{\alpha}_i\to \frac{1}{\gamma}{\bf c}_i^{\alpha}$ 
and with the substitution
$V(q)$ for its rational analog $\frac{1}{q}-\frac{1}{q+\gamma}$.
To present eqs.(\ref{qsK}) in the Lax form, 
in the paper \cite{KrZ} a spectral-dependent $L$-operator $L(z)$ was
suggested. One can see that in the rational case 
$L(z)$ coincides with $\bf L$ in the limit $z\to \infty$.
Thus, we obtain the Hamiltonian formulation of the spin generalization
of the rational RS model. 

Now we show how equations of motion (\ref{qd})-(\ref{cd}) can be solved
in terms of the factorization problem (see also \cite{OP,RS,RaS}). 
The Hamiltonian $H_R$ induces
on ${\cal P}$ equations of motion: 
$$ \dot{g}=0,~~~\dot{A}=g,~~~\dot{S}=0$$
that can be easily integrated: $A(t)=gt+A_0$, $g(t)=const$, $S(t)=const$. 
We  suppose that the positions of particles at $t=0$ are given by $q_i$ lying 
on ${\cal P}_r$.  It means that $A(t)=gt+Q$. Since for any $t$ the point 
$(A(t),g(t),S(t))$ satisfies the constraint (\ref{momf}) one gets
that $g$ should be identified with the $L$-operator $L_0$ at $t=0$.

Let us show that the solution of (\ref{qd}) is given by 
the diagonal factor $Q(t)$ in the decomposition of $A(t)$:
\be
A(t)=L_0t+Q=T(t)Q(t)T^{-1}(t),~~~T(t)e=e.
\la{eqm}
\ee
In \cite{AF} it was proved that 
\bea
\la{TdA}
\frac{\delta T_{ij}}{\delta A_{mn}}&=&\sum_{a\neq j}\frac{1}{q_{ja}}
(T_{ia}T_{nj}T_{am}^{-1}+T_{ij}T_{na}T_{jm}^{-1}),\\
\la{qdA}
\frac{\delta q_{i}}{\delta A_{mn}}&=&T_{ni}T_{im}^{-1}.
\eea
Using these formulae, we find 
$$
\dot{q}_i=\sum_{mn}\frac{\delta q_i}{\delta A_{mn}}\frac{dA_{mn}}{dt}=
(T^{-1}gT)_{ii}(t)=L_{ii}(t).
$$
Differentiating $\dot{q}_i$ once again, one gets 
\be
\ddot{q}_i=\dot{L}_{ii}=[T^{-1}gT,T^{-1}\dot{T}]_{ii},
\la{dqd}
\ee
where
\be
(T^{-1}\dot{T})_{ij}=-\frac{1}{q_{ij}}L_{ij}+\delta_{ij}\sum_{a\neq j}
\frac{1}{q_{ja}}L_{ja}.
\la{TdT}
\ee
Substituting eq.(\ref{TdT}) in eq.(\ref{dqd}), we obtain eq.(\ref{qd}).

As to the spin variables their equations of motion are automatically solved
if one knows the factor $T(t)$ in the decomposition (\ref{eqm}). Indeed,
if we define $\tilde{a}_i^{\alpha}(t)=(T^{-1}(t)a^{\alpha})_i$ then
$$
\dot{\tilde{a}}_i^{\alpha}=
-\sum_{j\neq i}
\frac{1}{q_{ij}}(\tilde{a}_i^{\alpha}-\tilde{a}_j^{\alpha})L_{ij} $$ 
and for the invariant spin 
${\bf a}_i^{\alpha}=\frac{1}{t_i}\tilde{a}_i^{\alpha}$ we get 
eq.(\ref{ad}). Solution of the equation of motion for ${\bf c}_i^{\alpha}$
is given by ${\bf c}_i^{\alpha}(t)=t_i(t)(b^{\alpha}L_0T(t))_i$.

The integrals of motion $J_n^{\alpha\beta}=\mbox{tr}(g^nS^{\alpha\beta})$ 
introduced in Section 2 take on ${\cal P}_r$ the following form
$$
J_n^{\alpha\beta}=\sum_{ij}({\bf L}^{n-1})_{ij}{\bf a}^{\alpha}_j
{\bf c}^{\beta}_i.
$$
Substituting here the explicit form (\ref{Li}) of the $L$-operator,
one can recast $J_n^{\alpha\beta}$ for $n\geq 1$ in the form
\bea
\nonumber
J_1^{\alpha\beta}&=&\sum_{i}S_i^{\beta\alpha},\\
J_n^{\alpha\beta}&=&\sum_{i_1,\dots,i_n}
\frac{(S_{i_1}S_{i_2}\dots S_{i_n})^{\beta\alpha}}
{(q_{i_1i_2}+\gamma)(q_{i_2i_3}+\gamma)\dots (q_{i_{n-1}i_n}+\gamma)},
\la{Int}
\eea
where we use $l\times l$-matrices
$S_i^{\alpha\beta}={\bf c}_i^{\alpha}{\bf a}_i^{\beta}$ ($i=1,\ldots N$).

An important property of  
$S_i$-variables is that they form a set of gauge-invariant variables
equivalent to $({\bf a},{\bf c})$. In fact, one can see that
$$
{\bf c}_i^{\alpha}=\sum_{\beta}S_i^{\alpha\beta}, ~~~
{\bf a}_i^{\alpha}=\frac{S_i^{\beta\alpha}}{\sum_{\gamma}S_i^{\beta\gamma}}.
$$
The Poisson structure of the model can be conveniently rewritten
in terms of $S_i^{\alpha\beta}$:
\bea
\nonumber
\{S_i^{\alpha\beta},S_j^{\mu\nu}\}&=&\frac{1}{q_{ij}}
(S_i^{\mu\beta}S_j^{\alpha\nu}+S_i^{\alpha\nu}S_j^{\mu\beta})
-\frac{\delta^{\beta\mu}}{q_{ij}+\gamma}(S_iS_j)^{\alpha\nu}
+\frac{\delta^{\alpha\nu}}{q_{ji}+\gamma}(S_jS_i)^{\mu\beta},\\
\{q_i,S_j^{\alpha\beta}\}&=&S_j^{\alpha\beta}\delta_{ij}.
\la{qS}
\eea
Since the Hamiltonian $H_R$ can be expressed as $H_R=\sum_i\mbox{Tr}S_i$,
eq.(\ref{qd}) acquires the form
\be 
\la{qs}
\ddot{q}_i=\frac{1}{\gamma^2}\sum_{j\neq i}
\mbox{Tr}(S_iS_j)(V(q_{ij})-V(q_{ji})),
\ee
where $\mbox{Tr}$ is used to denote the trace of an $l\times l$-matrix.
Analogously, eqs.(\ref{ad}) and (\ref{cd}) produce the equations of motion 
for $S_i$:
\be
\dot{S}_i=\frac{1}{\gamma}\sum_{j\neq i}(S_iS_jV(q_{ij})-S_jS_iV(q_{ji})).
\la{eqS}
\ee

Observe that the Poisson structure (\ref{qS}) and the Hamiltonian $H_R$
are invariant under the transformations $S_i\to \Omega^{-1}S_i\Omega$,
where $\Omega\in GL(l,{\bf C})$. These transformations are generated
by $J_0^{\alpha\beta}$.

Thus, we see that the Hamiltonian formalism of the rational spin RS model 
can be equivalently presented in terms of either $({\bf a},{\bf c})$- 
or $S_i$-variables.

The definition of ${\bf c}_i^{\alpha}=t_i(b^{\alpha}UP)_i$ implies 
that they contain the variables conjugated to $q_i$. However, we can 
not identify them with $P_i$ since the latter are not gauge-invariant.
The gauge-invariant momentum ${\bf P}_i$ can be defined as 
${\bf P}_i=u_i^{-1}P_it_i$. Computing the Poisson brackets 
of ${\bf P}_i$, one gets that $\{{\bf P}_i,{\bf P}_j\}=0$ and
$\{q_i,{\bf P}_j\}=\delta_{ij}{\bf P}_j$.

Recall that the invariant $L$-operator has the form 
${\bf L}=t^{-1}T^{-1}UPt$. Thus, it can be written as 
${\bf L}={\bf W}{\bf P}$, where ${\bf W}$ is 
a gauge-invariant variable: 
$$
{\bf W}=t^{-1}T^{-1}Uu.
$$
Then it is easy to see that ${\bf W}$ belongs to the Frobenius group, i.e.
it obeys the condition ${\bf W}e=e$:
\be
({\bf W}e)_i=\sum_{k,m}(t^{-1}T^{-1})_{ik}U_{km}(U^{-1}a^{\alpha})_m=
\frac{1}{t_i}(T^{-1}a^{\alpha})_i=1.
\la{We}
\ee
Just as it was for the spinless RS model, the Poisson bracket 
for ${\bf W}$ coincides with the Sklyanin bracket:  \be \{{\bf 
W}_1,{\bf W}_2\}=[r_{12},{\bf W}_1{\bf W}_2].  \la{WW} \ee

On ${\cal P}_r$ the variable ${\bf W}$ acquires the form
\be
{\bf W}_{ij}=\sum_{\alpha}\frac{{\bf 
a}_i^{\alpha}{\bf b}_j^{\alpha}}{q_{ij}+\gamma}, \la{Wr} 
\ee 
where 
${\bf b}_i^{\alpha}={\bf c}_i^{\alpha}{\bf P}_i^{-1}$. 

Remind that
the variables ${\bf a}_i^{\alpha}$ obey the constraints 
$\sum_{\alpha}{\bf a}_i^{\alpha}=1$ for any $i$. The condition 
(\ref{We}) implies that ${\bf b}_i^{\alpha}$ are also not arbitrary
but subject to the constraints:
\be
\sum_{\alpha}{\bf a}_i^{\alpha}
\sum_j\frac{{\bf b}_j^{\alpha}}{q_{ij}+\gamma}=1
\la{Wcon}
\ee
for any $i$. Therefore, the number of independent spin variables is 
$2N(l-1)$. In terms of these variables the Poisson structure of 
${\cal P}_r$ looks as follows:
\bea
\nonumber
\{q_i,{\bf P}_j\}&=&\delta_{ij}{\bf P}_j,~~~
\{q_i,{\bf a}_j^{\alpha}\}=0=\{q_i,{\bf b}_j^{\alpha}\},\\
\nonumber
\{{\bf P}_i,{\bf a}^{\alpha}_j\}&=&\frac{1}{q_{ij}}
({\bf a}^{\alpha}_i-{\bf a}^{\alpha}_j){\bf P}_i,\\
\nonumber
\{{\bf P}_i,{\bf b}^{\alpha}_j\}&=&(\delta_{ij}-{\bf W}_{ij} 
+\frac{1}{q_{ij}}{\bf b}^{\alpha}_j+
\delta_{ij}\sum _{n\neq i}\frac{1}{q_{nj}}
{\bf b}^{\alpha}_n ){\bf P}_i,\\ 
\nonumber
\{{\bf a}^{\alpha}_i,{\bf a}^{\beta}_j\}&=&
\frac{1}{q_{ij}}
({\bf a}^{\alpha}_i{\bf a}^{\beta}_j+{\bf a}^{\alpha}_j{\bf a}^{\beta}_i
-{\bf a}^{\alpha}_i{\bf a}^{\beta}_i-{\bf a}^{\alpha}_j{\bf a}^{\beta}_j),\\
\nonumber
\{{\bf b}_i^{\alpha},{\bf b}_j^{\beta}\}&=&
\delta_{ij}({\bf b}_i^{\beta}-{\bf b}_i^{\alpha})+\frac{1}{q_{ij}}
({\bf b}_j^{\alpha}{\bf b}_i^{\beta}-{\bf b}_j^{\beta}{\bf b}_i^{\alpha})
+\delta_{ij}\sum_{n\neq i}\frac{1}{q_{in}}
({\bf b}_n^{\beta}{\bf b}_i^{\alpha}-
{\bf b}_n^{\alpha}{\bf b}_i^{\beta}),\\
\la{ps}
\{{\bf a}_i^{\alpha},{\bf b}_j^{\beta}\}&=&-\delta^{\alpha\beta}
{\bf W}_{ij}+{\bf a}_i^{\alpha}{\bf W}_{ij},
\eea
where ${\bf W}$ is given by (\ref{Wr}).
One should point out that the structure (\ref{ps}) is Poisson only 
due to the constraints imposed on the spin variables. Therefore,
the rational spin RS model provides a new realization of the Poisson 
relations (\ref{WW}) as well as the $L$-operator algebra (\ref{L}).

Now we discuss the problem of generalizing the found Poisson structure
for the spin rational RS model to the trigonometric case.

Relaying on the fact that both in the spin and spinless cases 
the $L$-operator algebras may be the same, one can easily derive the 
trigonometric analog of the Poisson bracket (\ref{ff}) for the variables
$f_{ij}$. It follows from the results 
of \cite{AFM} that the trigonometric RS model can be described   
by the $L$-operator algebra (\ref{L}), where this time the $r$-matrices
$r$, $\bar{r}$ and $\hat{r}$ are given by
\bea
\la{rtr}
&&r=\sum_{ij}E_{ij}\otimes E_{ji}+
\sum_{i\neq j}\coth{(q_{ij})}E_{ii}\otimes E_{jj}+
\sum_{i\neq j}\coth{(q_{ij})}E_{ij}\otimes E_{ji}\\
\nonumber
&&~~~~~-
\sum_{i\neq j}\frac{e^{q_{ij}}}{\sinh{(q_{ij})}}E_{ij}\otimes E_{jj}+
\sum_{i\neq j}\frac{e^{q_{ij}}}{\sinh{(q_{ij})}}E_{jj}\otimes 
E_{ij},\\ \la{rchtr} 
&&
\bar{r}=-\sum_{i}E_{ii}\otimes E_{ii}+ 
\sum_{i\neq j}\coth{(q_{ij})}E_{ii}\otimes E_{jj}-
\sum_{i\neq j}\frac{e^{q_{ij}}}{\sinh{(q_{ij})}}E_{ij}\otimes 
E_{jj},~~~~~\\ 
\la{rftr} 
&&\hat{r}=-\sum_{ij}E_{ij}\otimes E_{ji}+ 
\sum_{i\neq j}\coth{(q_{ij})}(E_{ii}\otimes E_{jj}-E_{ij}\otimes 
E_{ji}).  
\eea 
For the spinless trigonometric RS model the 
$L$-operator satisfying (\ref{L}) is of the form $$ 
L=\sum_{ij}\frac{e^{q_{ij}+\gamma}}{\sinh{(q_{ij}+\gamma)}}c_j E_{ij},
$$
where $c_j$ are some functions of dynamical variables. It is natural
to assume that in the spin case the corresponding $L$-operator has
the form
$$
L=\sum_{ij}\frac{e^{q_{ij}+\gamma}}{\sinh{(q_{ij}+\gamma)}}f_{ij} 
E_{ij}.  
$$ 
Then the Poisson relations for $f_{ij}$ follow 
immediately from the $L$-operator algebra (\ref{L}):  \bea \la{trig} 
\{f_{ij},f_{kl}\}&=&
(\coth{(q_{ik})}+\coth{(q_{jl})}+\coth{(q_{kj})}+\coth{(q_{li})})f_{ij}f_{kl}
\\
\nonumber
&+&
(\coth{(q_{ik})}+\coth{(q_{jl})}+\coth{(q_{kj}+\gamma)}-\coth{(q_{il}+\gamma}))f_{il}f_{kj}
\\
\nonumber
&+&
(\coth{(q_{ki})}+\coth{(q_{il}+\gamma}))f_{ij}f_{il}
+(\coth{(q_{jk})}-\coth{(q_{jl}+\gamma)})f_{ij}f_{jl}\\
\nonumber
&+&
\nonumber
(\coth{(q_{ki})}-\coth{(q_{kj}+\gamma)})f_{kj}f_{kl}+
(\coth{(q_{il})}+\coth{(q_{lj}+\gamma)})f_{lj}f_{kl}
\eea
and they look like a trigonometric generalization of (\ref{ff}).

Now one can easily verify that equations of motion for $f_{ij}$
are given by (\ref{eqf}) with the potential 
$V(q)=\coth(q)-\coth(q+\gamma)$ and with the change of the overall 
factor $\frac{1}{\gamma}$ by $\frac{e^{\gamma}}{\sinh(\gamma)}$.  
On the other hand, these equations follow from eqs.(\ref{aK}) and 
(\ref{bK}). The problem of describing the Poisson structure of the 
trigonometric spin RS model would be completely solved if one finds 
such Poisson brackets for variables ${\bf a}$ and ${\bf c}$ that 
induce the ones (\ref{trig}) for $f_{ij}$. The straightforward 
generalization of eqs.(\ref{aa}-\ref{ac}) to the case at hand by 
replacing $\frac{1}{q_{ij}}$ by $\coth(q_{ij})$ is failed.  At the 
moment we can not offer a solution of the problem. 

\section{Euler-Calogero-Moser model}
We start this section with
discussing the degeneration of the spin RS system to the 
rational ECM model. For this purpose
we rescale $\log {\bf P}_i={\bf p}_i\to \varepsilon {\bf p}_i$ and $q_i\to 
\frac{1}{\varepsilon}q_i$, and consider the limit $\varepsilon\to 0$. 
The constraint (\ref{We}) implies that in this limit \be 
\sum_{\alpha}{\bf a}_i^{\alpha}{\bf b}_i^{\alpha}={\bf S}_{ii}=\gamma
\la{con}
\ee
and ${\bf W}_{ij}$ has the following expansion
$$
{\bf W}_{ij}=\delta_{ij}+\varepsilon(1-\delta_{ij})
\frac{{\bf S}_{ij}}{q_{ij}}-
\varepsilon \delta_{ij}\sum_{k\neq i}
\frac{{\bf S}_{ik}}{q_{ik}}+o(\varepsilon),
$$
where ${\bf S}_{ij}=\sum_{\alpha}{\bf a}_i^{\alpha}{\bf b}_j^{\alpha}$.
The corresponding expansion of the $L$-operator produces in the first 
order in $\varepsilon$ the $L$-operator ${\cal L}$ of the rational 
ECM model:  
\be 
{\cal L}_{ij}=\delta_{ij}(
{\bf p}_i- \sum_{k\neq 
i}\frac{{\bf S}_{ik}}{q_{ik}})+(1-\delta_{ij}) \frac{{\bf 
S}_{ij}}{q_{ij}}. 
\la{CL} 
\ee
In the limit $\varepsilon\to 0$ the Poisson structure (\ref{ps})
reduces to 
\bea
\nonumber
\{q_i,{\bf p}_j\}&=&\delta_{ij},~~~
\{q_i,{\bf a}_j^{\alpha}\}=0=\{q_i,{\bf b}_j^{\alpha}\},\\
\nonumber
\{{\bf p}_i,{\bf a}^{\alpha}_j\}&=&\frac{1}{q_{ij}}
({\bf a}^{\alpha}_i-{\bf a}^{\alpha}_j),\\
\nonumber
\{{\bf p}_i,{\bf b}^{\alpha}_j\}&=&
\delta_{ij}\sum_{k\neq i}\frac{{\bf S}_{ik}}{q_{ik}}-
(1-\delta_{ij})\frac{{\bf S}_{ij}}{q_{ij}}+ 
\frac{1}{q_{ij}}{\bf b}^{\alpha}_j+ \delta_{ij}\sum 
_{n\neq i}\frac{1}{q_{nj}} {\bf b}^{\alpha}_n ,\\ 
\nonumber 
\{{\bf a}^{\alpha}_i,{\bf a}^{\beta}_j\}&=& 0,\\ 
\nonumber \{{\bf b}_i^{\alpha},{\bf b}_j^{\beta}\}&=& \delta_{ij}({\bf 
b}_i^{\beta}-{\bf b}_i^{\alpha}),\\ \la{psdeg} \{{\bf 
a}_i^{\alpha},{\bf b}_j^{\beta}\}&=&-\delta^{\alpha\beta} 
\delta_{ij}+{\bf a}_i^{\alpha}\delta_{ij},
\eea
Introducing new momenta 
$$
p_i={\bf p}_i- \sum_{k\neq i}\frac{{\bf S}_{ik}}{q_{ik}},
$$
one can check that they have vanishing Poisson brackets with
${\bf a}^{\alpha}_i$ and ${\bf b}^{\alpha}_i$. The $L$-operator
${\cal L}$ turns into the standard one used in description of the ECM
system \cite{GH}-\cite{BAB}:
\be 
{\cal L}_{ij}=\delta_{ij}p_i+(1-\delta_{ij}) \frac{{\bf S}_{ij}}{q_{ij}}.
\la{CLst} 
\ee
To make a contact with the usual description of the ECM system
we introduce $lN$ pairs of canonical variables: 
$\{a_i^{\alpha},b_j^{\beta}\}=-\delta_{ij}\delta^{\alpha\beta}$.
Then the invariant variables ${\bf a}_i^{\alpha}$ and 
${\bf b}_j^{\beta}$ with the Poisson algebra (\ref{psdeg})
can be realized as 
$$
{\bf a}_i^{\alpha}=\frac{a_i^{\alpha}}{\sum_{\beta}a_i^{\beta}},~~~
{\bf b}_i^{\alpha}=b_i^{\alpha} \sum_{\beta}a_i^{\beta}.
$$

It is interesting to note that the Poisson algebra of the variables
${\bf S}_{ij}$ coincides with the defining relations of the 
Frobenius Lie algebra: 
\be 
\{{\bf S}_{ij},{\bf 
S}_{kl}\}=\delta_{il}({\bf S}_{ij}-{\bf S}_{kj})+ \delta_{jl}({\bf 
S}_{kj}-{\bf S}_{ij})+ \delta_{jk}({\bf S}_{il}-{\bf S}_{kl}).  
\la{Sin}
\ee
These relations are compatible with the constraint 
${\bf S}_{ii}=\gamma$. 

The appearance of the Frobenius spin variables in the rational
ECM model is not accidental. In fact, the same phenomenon takes
place for the general elliptic ECM model. To elucidate this fact
we recall that the elliptic ECM system is described by the Hamiltonian 
$H=\frac{1}{2}\sum_ip_i^2-\frac{1}{2}\sum_{i\neq j}S_{ij}S_{ji}V(q_{ij})$,
where $V(q_{ij})={\cal P}(x)$ is the Weierstrass ${\cal P}$-function 
and $S_{ij}$ are the spin variables defined by (\ref{Ss}) and having
the Poisson bracket:
$$
\{S_{ij},S_{kl}\}=\delta_{jk}S_{il}-\delta_{il}S_{kj}.
$$ 
The model is described by the $L$-operator \cite{BAB}
$$
L=\sum_{i}(p_i+\zeta(z)S_{ii})E_{ii}+\sum_{i\neq 
j}\Phi(z,q_{ij})S_{ij}E_{ij}, 
$$ 
where 
$\Phi(z,q)=\frac{\sigma(z+q)}{\sigma(z)\sigma(q)}$. 
This $L$-operator satisfies the 
Poisson algebra 
\bea  
\nonumber
\{L_1(z),L_2(w)\}&=&[r_{12}(z,w),L_1(z)]-[r_{21}(w,z),L_2(w)]\\
\la{Lc}
&+&\sum_{i\neq j}\frac{\partial}{\partial q_{ij}}\Phi(z-w,q_{ij})
(S_{ii}-S_{jj})E_{ij}\otimes E_{ji}, 
\eea
with the dynamical $r$-matrix
\be
r_{12}(z,w)=\zeta(z-w)\sum_{i}E_{ii}\otimes E_{ii}+
\sum_{i\neq j}\Phi(z-w,q_{ij}) E_{ij}\otimes E_{ji}.
\la{dr}
\ee
Due to the last term in (\ref{Lc}) the model is not integrable. However,
the Hamiltonian is invariant under the symmetry 
$a_i\to k_ia_i$, $b_i\to \frac{1}{k_i}b_i$ generated by $S_{ii}$.
The integrability is obtained on the reduced space $S_{ii}=const$.
As in the case of the RS system, to perform the reduction we define
the gauge-invariant $L$-operator ${\bf L}=tLt^{-1}$ with 
$t_{ij}=\delta_{ij}\sum_{\alpha}a^{\alpha}_i$ or explicitly
\be
{\bf L}=\sum_{i}(p_i+\zeta(z))E_{ii}+
\sum_{i\neq j}\Phi(z,q_{ij}){\bf S}_{ij}E_{ij},
\la{Cin}
\ee
where the gauge-invariant spin variables appeared 
$$
{\bf S}_{ij}=S_{ij}\frac{\sum_{\alpha}a^{\alpha}_j}
{\sum_{\alpha}a^{\alpha}_i}.
$$
Computing the Poisson bracket of ${\bf S}_{ij}$, we obtain that it
precisely coincides with (\ref{Sin}).

Now it is easy to establish that the Poisson 
algebra of the $L$-operator (\ref{Cin}) has the form
$$
\{{\bf L}_1(z),{\bf L}_2(w)\}=[{\bf r}_{12}(z,w),{\bf L}_1(z)]-[{\bf 
r}_{21}(w,z), {\bf L}_2(w)], 
$$
where a matrix ${\bf r}$ literally coincides 
with the $r$-matrix of the elliptic Calogero-Moser model \cite{Skl,BS}:
\bea
\nonumber
{\bf r}_{12}(z,w)&=&(\zeta(z-w)+\zeta(w))\sum_{i}E_{ii}\otimes E_{ii}\\
\nonumber
&+&
\sum_{i\neq j}\Phi(z-w,q_{ij}) E_{ij}\otimes E_{ji}+
\sum_{i\neq j}\Phi(w,q_{ij}) E_{jj}\otimes E_{ij}.
\eea
Thus, the ECM model corresponds to a representation of 
the $L$-operator algebra of the Calogero-Moser model that depends 
not only on $q_i$ and $p_i$ but also on the additional spin variables
${\bf S}_{ij}$ with the bracket (\ref{Sin}). As to the 
spectral-dependent $L$-operator of the spin RS model, it
does not satisfy the $L$-operator algebra found for the 
spinless case \cite{NKSR,Sur1}. This algebra is 
quadratic what fixes the form of the corresponding $L$-operator
almost uniquely. 

\section{Conclusion}
In this paper we presented a detailed description of the Poisson 
structure for the rational spin RS model by using the Hamiltonian 
reduction technique. The results obtained can not be extended
to the trigonometric spin RS model in a straightforward manner.
It was shown in \cite{GNe} that the trigonometric RS model
can be obtained by means of the Poisson reduction technique 
applied to the Heisenberg double $D$ associated to $G=GL(N,{\bf C})$. 
Therefore, one may hope to describe the Poisson structure of
the trigonometric spin RS model in the same fashion starting
from the phase space $D\times G^*$, where $G^*$ is a Poisson-Lie 
group dual to $G$. 

As is known \cite{GH} the rational ECM model possesses the current 
algebra symmetry and, as we have established, the  same
symmetry occurs in the rational spin RS model. On the other hand,
the trigonometric ECM model has the Yangian symmetry. Thus, 
for the trigonometric spin RS model it is natural to expect the 
appearance of the Yangian symmetry.

The elliptic case is much more involved since at the moment
a reduction procedure leading to the elliptic spin RS model
remains to be unknown.

Another interesting open problem is to quantize the spin RS models.
In the rational case one could use the quantum Hamiltonian reduction
procedure developed in \cite{AF}.

{\bf ACKNOWLEDGMENT} The authors are grateful to 
A.Zabrodin for valuable discussions. This work is supported in part by 
the RFBI grants N96-01-00608, N96-01-00551 and by the ISF 
grant a96-1516.  

\end{document}